\documentclass[aps,a4paper,showkeys,manuscript]{revtex4}
\usepackage{amsmath}
\usepackage{color}%
\usepackage{graphicx}
\usepackage{dcolumn}
\usepackage{bm}
\usepackage{booktabs}            
\usepackage{amssymb,bm,mathrsfs,bbm,amscd} 
\usepackage{longtable}

\begin{document}
\title{Fast neutron scattering on Gallium target at 14.8 MeV}
\author{R. Han$^{1,2,4}$, R. Wada$^{1}$, Z. Chen$^{1}$\footnote{zqchen@impcas.ac.cn},
Y. Nie$^{3}$, X. Liu$^{1,4}$, S. Zhang$^{1,4}$, P. Ren$^{1,4}$, B.
Jia$^{1,4}$, G. Tian$^{1,4}$, F. Luo$^{1,4}$, W. Lin$^{1,4}$, J.
Liu$^{1}$, F. Shi$^{1}$,  M. Huang$^{1}$, X. Ruan$^{3}$, J.
Ren$^{3}$, Z. Zhou$^{3}$, H. Huang$^{3}$, J. Bao$^{3}$, K.
Zhang$^{3}$, B. Hu$^{2}$}
\affiliation{$^{1}$Institute of Modern Physics, Chinese Academy of
Sciences, Lanzhou, 730000, China} \affiliation{$^{2}$School of
Nuclear Science and Technology, Lanzhou University, Lanzhou
730000, China} \affiliation{$^{3}$China Institute of Atomic
Energy, Beijing, 102413, China} \affiliation{$^{4}$University of
Chinese Academy of Sciences, Beijing, 100049, China}
\begin{abstract}
Benchmarking of evaluated nuclear data libraries was performed for
$\sim 14.8$ MeV neutrons on Gallium targets. The experiments were
performed at China Institute of Atomic Energy(CIAE). Solid samples
of natural Gallium (3.2 cm and 6.4 cm thick) were bombarded by
$\sim 14.8$ MeV neutrons and leakage neutron energy spectra were
measured at 60$^{\circ}$ and 120$^{\circ}$. The measured spectra
are rather well reproduced by MCNP-4C simulations with the
CENDL-3.1, ENDF/B-VII and JENDL-4.0 evaluated nuclear data
libraries, except for the inelastic contributions around $E_{n} =
10-13$ MeV. All three libraries significantly underestimate the
inelastic contributions. The inelastic contributions are further
studied, using the Talys simulation code and the experimental
spectra are reproduced reasonably well in the whole energy range
by the Talys calculation, including the inelastic contributions.
\end{abstract}

\maketitle
\section*{I. Introduction}
The experimental studies of fast neutron scattering are important
for design of nuclear reactors~\cite{Hiroyuki1986,Steven2012}.
They play a crucial role for verification of the evaluated nuclear
data libraries, especially the elements that are of interest in
accelerator driven systems, fission and fusion reactor
technologies. Gallium (Ga) is one of such elements, which can be
used as a cooling agent. Due to special physical properties of Ga,
it is also used to make filling materials of some element samples,
such as, Pu element within the device of reactors and nuclear
weapons in a mixed form with Ga. Ga is a soft silvery metal and
becomes liquid at slightly above the room temperature (29.78
$^{\circ}$C). It is a moderate heat conductor similar to Lead and
non-explosive unlike Sodium. Therefore it is a good candidate of a
cooling agent for nuclear reactors. However, its available
experimental data are limited and benchmarking of the evaluated
nuclear data libraries is necessary. The benchmarking has been
performed at CIAE since 2009~\cite{Nie2010}. The validity of the
benchmarking test system has been examined. Several targets have
been studied, such as U~\cite{Nie2010}, Be~\cite{Nie2013},
water~\cite{Zhang2014} and polyethylene~\cite{Nie2012}.

In this article, we studied fast neutron scatterings on Ga
samples. The leakage neutron energy spectra were measured by time
of flight measurements and evaluated nuclear data libraries were
benchmarked. The discrepancies between the experimental data and
the simulations based on the evaluated nuclear libraries are
further investigated. The article is organized as follows. In
Sec.II, the experiment is briefly described. In Sec.III, the
benchmarking of evaluated nuclear data libraries are presented. In
Sec.IV, elastic and inelastic contributions are discussed. A
summary is given in Sec.V.
\section*{II. Experiment}
\begin{figure}[htbp]
\includegraphics[width=8cm]{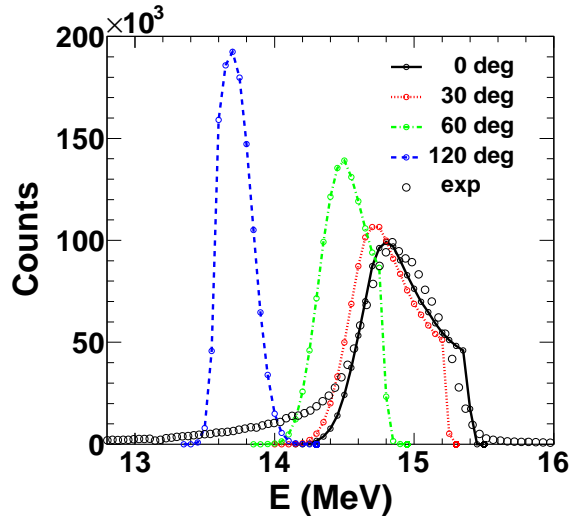}
\centering \caption{Energy distributions of the neutrons emitted
from the $T (d, n) ^{4}He$ source in different angles
($0^{\circ}$, $30^{\circ}$, $60^{\circ}$ and $120^{\circ}$) with
respect to the $D^{+}$ beam direction predicted by TARGET code.
Open squares are the experimental distributions from the
$0^{\circ}$ monitor detector, which is normalized to the
calculation at the peak value.} \label{fig:fig1}
\end{figure}

The experiment was performed at CIAE. Neutrons were produced by
the $T (d, n) ^{4}He$ reaction using the 300 keV $D^{+}$ beam
accelerated by the Cockcroft-Walton accelerator. The average beam
current was about 30 $\mu$A during the experiment. A silicon
surface barrier detector positioned at $135^{\circ}$ with respect
to the $D^{+}$ beam was used to monitor the neutron yield by
counting the associated $^{4}He$ particles. Another monitor, a
BC501A scintillation detector, was placed at 7.98 m from the
neutron source at approximately zero degree. The angles of the
leakage neutrons are selected by changing the target position
along the leakage neutron direction, which results in change of
the neutron emission angle from the Tritium target and the flight
distance of the target sample to the detector. The energy
distributions of the neutrons emitted from the $T (d, n) ^{4}He$
source in different angles, $0^{\circ}$, $30^{\circ}$,
$60^{\circ}$ and $120^{\circ}$ with respect to the $D^{+}$ beam
direction, have been simulated using TARGET
code~\cite{schlegel2005}. The calculated results are shown in
Fig.~\ref{fig:fig1} with the experimentally observed distribution
by the monitor scintillator near 0$^{\circ}$. The energy
distributions of the neutrons are around 14 MeV, slightly
depending on different emission angles. The average neutron energy
decreases slightly as the emission angle increases. The energy
distribution measured by the monitor detector is well reproduced
around the peak energy by the calculation at 0$^{\circ}$, which is
slightly affected by the accelerator pulse shapes. The pulse width
of the $D^{+}$ beam during this experiment was kept $\sim$ 3 ns.
In the experiment, the generated neutrons from two angles,
$-30^{\circ}$ and $30^{\circ}$ with respect to the $D^{+}$ beam
direction were used, which gives the leakage neutron emission
angle of 60$^{\circ}$ and 120$^{\circ}$, respectively, and the
same average incident neutron energy distribution peaked at around
14.8 MeV at both target positions. The beam was impinged on Ga
samples. Natural Ga was used, which is composed of $60.11\%$
$^{69}Ga$ and $39.89 \%$ $^{71}Ga$. Two self-supported solid Ga
samples were made as a cylindrical shape with $\phi13$ cm $\times$
3.2 cm and $\phi13$ cm $\times$ 6.4 cm. (The samples were
manufactured at a low temperature room and stored in a
refrigerator. During the experiment, room temperature was kept
$\sim 20^{\circ}$ to keep them in solid.) Leakage neutron spectra
from Gallium samples ware measured using a BC501A ($\phi5.08$ cm
$\times$ 2.54 cm) scintillation detector by a TOF technique with
the flight path of 8.30 m and 8.67 m for 60$^{\circ}$ and
120$^{\circ}$, respectively. A pre-collimator system, which is
made of Iron, polyethylene and Lead, were orderly placed along the
fight path between the sample and the detector. In front of the
detector, another collimator was embedded inside the concrete wall
of 2 m thick with a hole of $\sim 10$ cm diameter , which was set
to shield the neutron detector from background neutrons. Using
such a heavy shielding and collimating system, high
foreground/background ratio has been achieved. The light output
function and the detection efficiency of the neutron detector were
well calibrated at the 1-20 MeV at the Tandem Accelerator at
CIAE~\cite{Huang2009}. For a given thresholds, the detection
efficiency can be well determined with NEFF
code~\cite{Dietze1982}. Further details of the experimental setup
and the data acquisition system can be found in
Ref.~\cite{Nie2010} and~\cite{Nie2012}.

\section*{III. Results}
\begin{figure}[htbp]
\includegraphics[width=14cm]{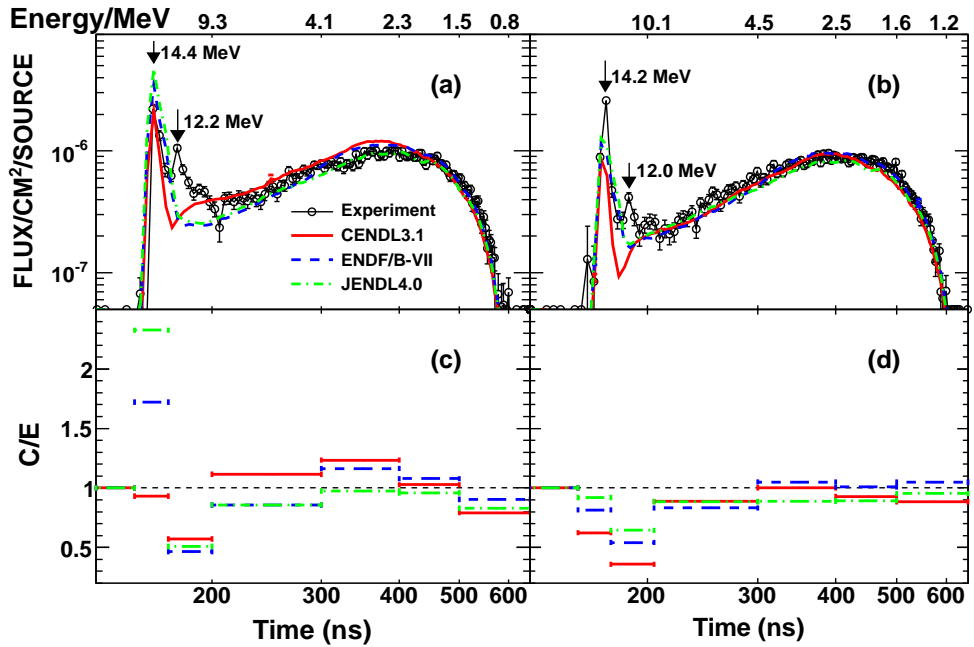}
\centering \caption{Neutron leakage time spectra measured in the
experiment and simulated by MCNP code at (a) $60^{\circ}$ (b)
$120^{\circ}$ for 6.4-cm-thick gallium sample; Circles connected
by line represent the experimental results, whereas different
lines are those from the MCNP simulations using the evaluated
databases of CENDL3.1 (solid), ENDF/B-VII (dashed) and JENDL4.0
(dot-dashed). The errors presented in the figure are from the
statistical errors. (c) and (d) $C/E$ ratios for
CENDL3.1/experiment, ENDF/B-VII/experiment and
JENDL4.0/experiment.} \label{fig:fig2}
\end{figure}

The neutron leakage time spectra measured in the experiment are
shown by symbols in Fig.~\ref{fig:fig2} (a) and (b). The measured
data were normalized to the n-p scattering cross section on a
polyethylene target in a separate run. Time calibration has been
made, using the gamma peak which is eliminated in
Fig.~\ref{fig:fig2}. The errors presented in the figure are from
the statistical errors. As shown in the figure, the neutron
leakage time spectra are quite similar with each other at both
angles. Similar characteristic properties are also observed for
those of the 3.2 cm thinner sample, and therefore their spectra
are not shown. In the highest energy region, a shape peak is
observed, which corresponds to the elastic scattering. The
experimentally observed widths of the elastic peaks are dominated
by the incident neutron energy distribution at the target which is
shown in Fig.1. Inelastic contributions start right after the
elastic peak. $(n, n'), (n, 2n), (n, np)$ and $(n, n\alpha)$
reaction channels are opened at the incident neutron energy of
14.8 MeV. These contributions are observed as a broad peak
starting around 170 ns in the time spectra (below 13 MeV in
energy).

To benchmark evaluated nuclear data libraries, the neutron leakage
spectra are simulated by MCNP-4C~\cite{Briesmeister} code using
Gallium evaluated nuclear data from the CENDL3.1, ENDF/B-VII and
JENDL4.0 libraries. In the MCNP simulations, the experimental
parameters are taken into account. These include the incident
neutron energy distribution at the target, the target thickness,
the neutron detection efficiency, the time response of the neutron
detector and among others. The calculated results of the MCNP
simulations are compared with those of the experiment in
Fig.~\ref{fig:fig2} (a) and (b). The general trends of the
experimental neutron leakage spectra are well reproduced except
for the inelastic peak around $E_n \sim 12$ MeV.

In order to make further detail comparisons, the ratios of
calculated and experimental cross sections, $C/E$, are plotted in
Fig. 2 (c) and (d) as a function of time for a certain time
interval. From the above comparisons, we made the following
observations:

\begin{enumerate}
\item{At $t = 150-170$ ns ($E_n \sim 14$ MeV), a sharp elastic
peak is observed. The simulated yield of the elastic scattering
neutrons depends on the choice of the evaluated nuclear libraries.
For $60^{\circ}$, the yield from the CENDL3.1 library agrees
within 5$\%$ comparing with that of the experiment, whereas those
from the ENDF/B-VII and JENDL4.0 libraries are larger by a factor
of $\sim 1.8$ and a factor of $\sim 2.4$, respectively. At
$120^{\circ}$, the ratios are 0.6, 0.8, and 0.9 for the CEND3.1,
ENDF/B-VII, and JENDL4.0, respectively.}

\item{At $t = 170-200$ ns ($E_n \sim 10-13$ MeV), a small, but
clear peak is observed. The contribution originates from inelastic
scatterings and all MCNP simulations using these three libraries
do not show a peak as the experimental data do and significantly
underestimate the cross section.}

\item{At $t > 200$ ns ($E_n < 10$ MeV), a broad peak is observed,
which originates from the $(n, n'), (n, 2n), (n, np)$ and $(n,
n\alpha)$ reaction channels. The experimental cross sections are
well reproduced by the simulations from all three libraries within
20$\%$ at both angles.}
\end{enumerate}

\section*{IV. Discussions}
In the following discussions, we first address the discrepancy of
the elastic contribution for the different nuclear libraries. Then
we address the problem of the inelastic contributions, using the
Talys program.
\section*{A. Elastic contribution}
\begin{figure}[htbp]
\includegraphics[width=14cm]{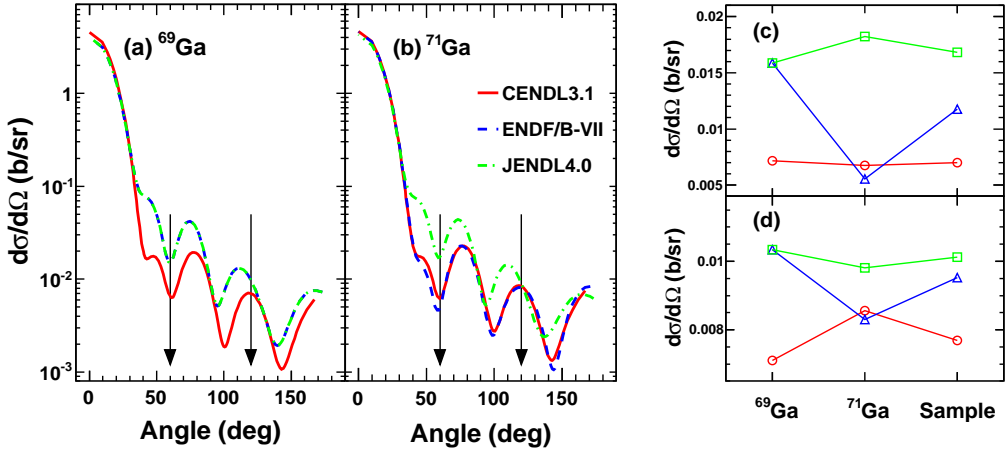}
\centering \caption{Evaluated angular distributions of the elastic
scattering channels for $^{69}Ga$ (a) and $^{71}Ga$ (b) given in
the libraries of CENDL3.1, ENDF/B-VII and JENDL4.0. Different
lines indicate different evaluated data from different libraries.
Evaluated $60^{\circ}$ (c) and $120^{\circ}$ (d) elastic cross
sections of $^{69}Ga$, $^{71}Ga$ and natural sample from
Fig.~\ref{fig:fig3} (a) and (b) respectively.} \label{fig:fig3}
\end{figure}

The angular distributions of the elastic scattering for $^{69}Ga$
and $^{71}Ga$, from the CENDL3.1, ENDF/B-VII and JENDL4.0
libraries, are shown in Fig.~\ref{fig:fig3} (a) and (b),
respectively. The available data of the three evaluated nuclear
libraries at 14.75 MeV are used for the plots. As shown in
Fig.~\ref{fig:fig3} (a) and (b), the angular distributions show a
strong forward peaking and an oscillation pattern appears at
angles larger than $30^{\circ}$. The oscillation patterns of the
evaluated elastic angular distributions of $^{69}Ga$ and $^{71}Ga$
from the CENDL3.1 and JENDL4.0 libraries are almost identical,
whereas those from the ENDF/B-VII library show slightly different
distributions between $^{69}Ga$ and $^{71}Ga$. The elastic cross
sections of $^{69}Ga$ and $^{71}Ga$ at $60^{\circ}$ and
$120^{\circ}$, are picked up from Fig.~\ref{fig:fig3} (a) and (b),
at angle positions indicated by arrows, and plotted in
Fig.~\ref{fig:fig3} (c) for $60^{\circ}$ and (d) for $120^{\circ}$
with the calculated values for the natural Ga sample. One should
note that $60^{\circ}$ in the angular distribution locates near
the valley of the distributions for all cases, and thus the effect
of the ambiguity of the experimental angular determinations on the
cross section becomes an order of less than 10$\%$. $120^{\circ}$,
on the other hand, locates near a peak for some cases. In this
case the ambiguity from the angular determination becomes in the
similar order, whereas for other cases, the angle locates on the
shoulder of the distribution and thus the ambiguity becomes
larger. The cross section of the natural target sample is
calculated as
$\sigma=0.6011\sigma(^{69}Ga)+0.3989\sigma(^{71}Ga)$. The ratios
of the elastic cross sections of the target sample between these
three libraries, are $\sigma _{CENDL3.1}:\sigma
_{ENDF/B-VII}:\sigma _{JENDL4.0} \sim 1.0:1.7:2.4$ at $60^{\circ}$
and $\sim 0.7:0.9:1.0$ at $120^{\circ}$. These ratios are quite
consistent with those obtained in Fig.~\ref{fig:fig2} (c) and (d),
in which these ratios of $1.0:1.8:2.4$ at $60^{\circ}$ and
$0.6:0.8:0.9$ at $120^{\circ}$ are obtained. These observations
indicate that the discrepancies of the elastic scattering neutron
yields in the MCNP simulations using the CENDL3.1, ENDF/B-VII and
JENDL4.0 libraries in Fig.~\ref{fig:fig2} (a) and (b) originate
simply from the differences in the angular distribution of the
elastic cross sections at $60^{\circ}$ and $120^{\circ}$ in these
libraries. One should note that, as shown in Fig.~\ref{fig:fig3}
(a) and (b), the major contribution of the elastic scattering to
the total reaction cross section is governed by the contribution
at angles smaller than $30^{\circ}$ and therefore the
discrepancies in the elastic peaks observed in this experiment is
in a minor effect for the whole elastic contribution.

\section*{B. Inelastic contributions}
\begin{figure}[htbp]
\includegraphics[width=10cm]{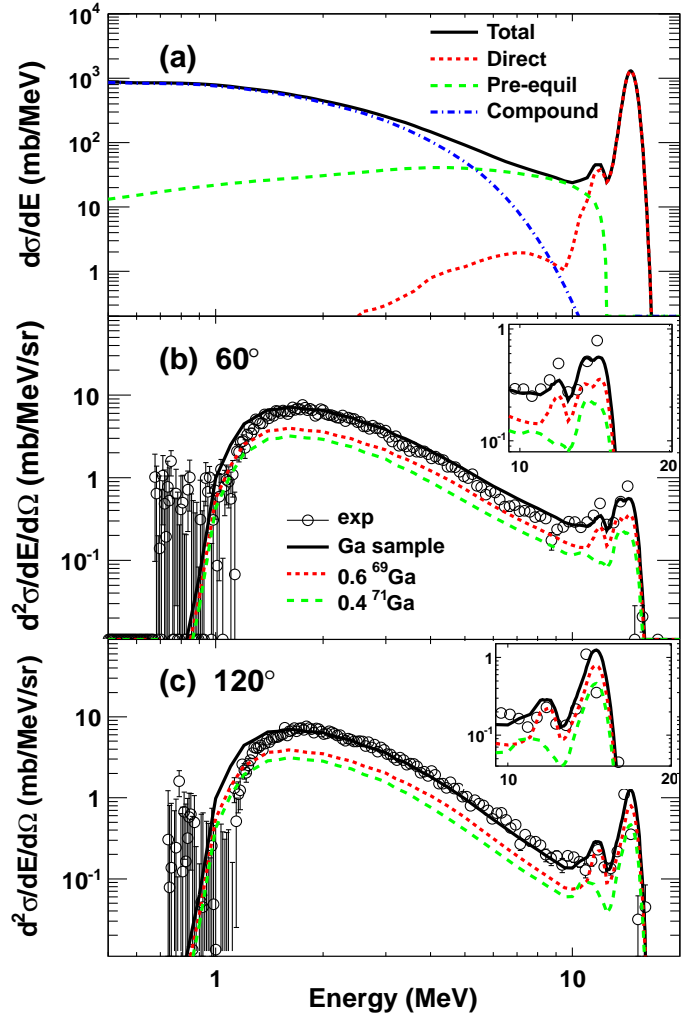}
\centering \caption{(a) The total neutron leakage energy spectrum
and the contributions from three processes (the direct,
pre-equilibrium and compound processes) calculated by Talys code.
The experimental spectra at $60^{\circ}$ (b) and $120^{\circ}$ (c)
are compared with the Talys calculated for natural Ga, $^{69}$Ga
and $^{71}$Ga, respectively. The detector efficiency has been
considered in the Talys calculation. Insets show expanded energy
spectra above 10 MeV.} \label{fig:fig4}
\end{figure}

To investigate the discrepancies in the neutron leakage energy
spectra between the experiment and the MCNP simulations at $E_n
\sim 10-13$ MeV, simulations by Talys-1.6 code~\cite{koning2013}
are performed. In the Talys code, the neutron-induced reaction is
divided in three physical processes, e.g., direct, pre-equilibrium
and compound nucleus reactions. Calculated total neutron leakage
energy spectrum and the contributions from three processes are
shown in Fig.~\ref{fig:fig4} (a). Natural Ga samples was used in
the plots. The calculated total cross section show three
\begin{figure}[htbp]
\includegraphics[width=10cm]{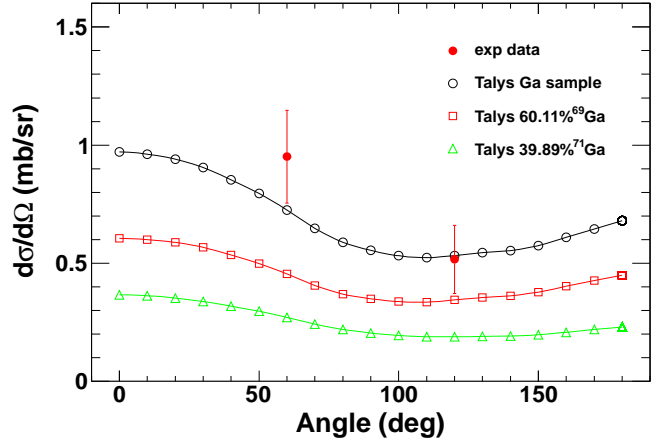}
\centering \caption{Angular distribution the inelastic cross
section integrated over $E_n = 10-12.5$ MeV from Talys
simulations. The red dots are the experimental data for
$60^{\circ}$ and $120^{\circ}$.} \label{fig:fig5}
\end{figure}
\begin{figure}[htbp]
\includegraphics[width=10cm]{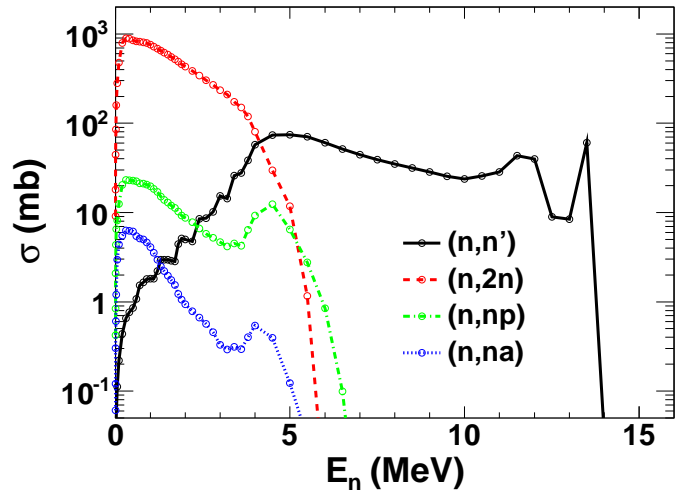}
\centering \caption{The contributions of inelastic reaction from
$(n,2n), (n,np), (n,na)$ channels and a emission neutron in the
continuum states ($0 - 12$ MeV) and discrete levels ($12 - 14$
MeV) from $(n,n')$ channel in natural Ga sample. These results
calculated by Talys code.} \label{fig:fig6}
\end{figure}
\begin{figure}[htbp]
\includegraphics[width=8cm]{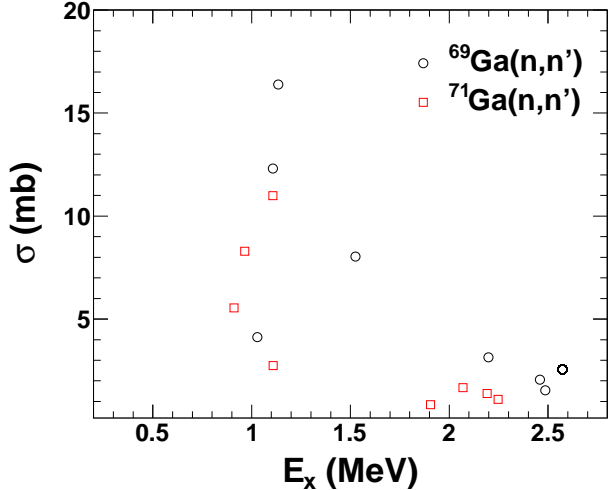}
\centering \caption{Inelastic cross section of the leakage neutron
from $(n,n')$ channel in $^{69}$Ga (circles) and $^{71}$Ga
(squares) as a function of $E_{x}$.} \label{fig:fig7}
\end{figure}
characteristic features, an elastic peak, an inelastic peak and a
broad bump in the low energy side. The elastic and inelastic peaks
are generated by the direct process. The broad bump at the low
energy is dominated by the compound nucleus process. The
contribution from the pre-equilibrium process is dominated at the
energy range of $E_n \sim 5-10$ MeV. In Fig.~\ref{fig:fig4} (b)
and (c), the experimental energy spectra at $60^{\circ}$ and
$120^{\circ}$ are compared with the calculations, respectively.
The experimental spectra are well reproduced by the calculated
cross sections for the Ga sample, including the inelastic peaks at
both angles. Individual contributions of $^{69}$Ga and $^{71}$Ga
to those of the Ga sample are also shown by dotted and dashed
lines. As shown clearly in the inserts, the inelastic peaks are
mainly generated from the direct process of $^{69}$Ga, but not of
$^{71}$Ga though a very small contribution from $^{71}$Ga is
observed at $120^{\circ}$ at a slightly ($\sim1$ MeV) lower
neutron energy. In Fig.~\ref{fig:fig5} the calculated angular
distribution of the inelastic cross section at $10. 0 < E_n <
12.5$ MeV  are plotted from the Talys simulation (open symbols)
and compared with those of the experiment (filled circles). In
contrast to the elastic channel, the calculated angular
distribution is rather flat and the experimental data agree within
the error bars at both angles. In this work, since we are focusing
on the neutron emission channels, no neutron pick-up reaction
channel is used in the direct reaction channel. In
Fig.~\ref{fig:fig6}, the contributions of different reaction
channels to the total reaction cross sections are plotted as a
function of the leakage neutron energy for natural Ga sample. The
$(n, n')$ cross section is the sum of those from the discrete and
continuum states calculations. The $(n, n')$ channel dominates in
the direct and pre-equilibrium contribution at 5 MeV $\leq E_n
\leq 12$ MeV. At $E_n < 5$ MeV, the compound nucleus contribution
of the $(n,2n)$ process is dominated and $(n,np)$ and
$(n,n\alpha)$ channels show minor contributions. In
Fig.~\ref{fig:fig7}, the calculated cross section of the
individual state below 2.6 MeV is plotted for $^{69}$Ga(circles)
and $^{71}$Ga(squares). As one can see, the location of the
excited levels are similar for the two isotopes, but the strengths
are larger by about $50\%$ for $^{69}$Ga. For the natural Ga
samples, the natural abundance of $40\%$ is multiplied for
$^{71}$Ga and the relative yield to those of $^{69}$Ga is further
reduced. This causes a relatively small contribution from
$^{71}$Ga in the inelastic contribution at $E_n = 10-12$ MeV. This
fact indicates that there is no essential difference between the
reaction mechanisms of neutron scattering between $^{69}$Ga and
$^{71}$Ga in the inelastic channels.

\section*{V. summary}
The neutron leakage spectra from natural Ga samples at the
incident energy of $\sim$ 14.8 MeV are compared with the MNCP
simulations with the CENDL3.0, ENDF/B-VII and JENDL4.0 evaluated
nuclear data libraries at $60^{\circ}$ and $120^{\circ}$. The
essential characteristic properties of the spectra are well
reproduce by these simulations, except for the inelastic
contributions at $E_n \sim 12$ MeV. The results from all three
libraries significantly underestimate the cross section in this
energy range. The experimental spectra are further compared with
those of the Talys simulations with the standard default
parameters in addition to the parameters related to the
experiments. The Talys simulations reproduce the experimental
spectra very well, including the inelastic peak at  $E_n \sim 12$
MeV. It is found that the inelastic contributions originated
mainly from the $(n,n')$ channel of $^{69}$Ga and minor
contribution from those of $^{71}$Ga. As far as concerning to the
observed spectra for Ga samples in this experiment, the Talys code
provides a slightly better nuclear data information, comparing to
the CENDL3.0, ENDF/B-VII and JENDL4.0 evaluated nuclear data
libraries.

\section*{Acknowledgments}

The authors thank to the operational staff in the Cockcroft-Walton
accelerator, China Institute of Atomic Energy, for their support
during the experiment. The experiment is supported by the National
Natural Science Foundation of China (Grants No. 11075189) and ADS
project 302 (Grants No. XDA03030200) of the Chinese Academy of
Sciences.


\end{document}